\documentclass[sigconf,nonacm]{acmart}
\usepackage{Definitions}
\usepackage{algorithm}
\usepackage{algorithmic}

\usepackage{xcolor}
\newcommand{\given}{\,|\,}
\newcommand{\lgiven}{\,\bigg\rvert\,}

\AtBeginDocument{%
  \providecommand\BibTeX{{%
    \normalfont B\kern-0.5em{\scshape i\kern-0.25em b}\kern-0.8em\TeX}}}

\setcopyright{acmcopyright}
\copyrightyear{2023}
\acmYear{2023}
\setcopyright{acmlicensed}\acmConference[WSDM '23]{Proceedings of the Sixteenth ACM International Conference on Web Search and Data Mining}{February 27-March 3, 2023}{Singapore, Singapore}
\acmBooktitle{Proceedings of the Sixteenth ACM International Conference on Web Search and Data Mining (WSDM '23), February 27-March 3, 2023, Singapore, Singapore}
\acmPrice{15.00}
\acmDOI{10.1145/3539597.3570469}
\acmISBN{978-1-4503-9407-9/23/02}

\begin{document}

\title{Uncertainty Quantification for Fairness in Two-Stage Recommender Systems}

\author{Lequn Wang}
\affiliation{%
  \institution{Cornell University}
  \city{Ithaca, NY}
  \country{USA}}
\email{lw633@cornell.edu}

\author{Thorsten Joachims}
\affiliation{%
  \institution{Cornell University}
  \city{Ithaca, NY}
  \country{USA}}
\email{tj@cs.cornell.edu}


\begin{abstract}
Many large-scale recommender systems consist of two stages. The first stage efficiently screens the complete pool of items for a small subset of promising candidates, from which the second-stage model curates the final recommendations. In this paper, we investigate how to ensure group fairness to the items in this two-stage architecture. In particular, we find that existing first-stage recommenders might select an irrecoverably unfair set of candidates such that there is no hope for the second-stage recommender to deliver fair recommendations. To this end, motivated by recent advances in uncertainty quantification, we propose two threshold-policy selection rules that can provide distribution-free and finite-sample guarantees on fairness in first-stage recommenders. More concretely, given any relevance model of queries and items and a point-wise lower confidence bound on the expected number of relevant items for each threshold-policy, the two rules find near-optimal sets of candidates that contain enough relevant items in expectation from each group of items. To instantiate the rules, we demonstrate how to derive such confidence bounds from potentially partial and biased user feedback data, which are abundant in many large-scale recommender systems. In addition, we provide both finite-sample and asymptotic analyses of how close the two threshold selection rules are to the optimal thresholds. Beyond this theoretical analysis, we show empirically that these two rules can consistently select enough relevant items from each group while minimizing the size of the candidate sets for a wide range of settings. 
\end{abstract}


\maketitle
\section{Introduction}
Two-stage pipelines~\cite{covington2016deep,yi2019sampling,zhao2019recommending,chen2019top,he2014practical,borisyuk2016casmos,zhu2018learning,eksombatchai2018pixie} are ubiquitous in large-scale recommender systems. Their key advantage lies in their efficiency and scalability, making it possible to curate personalized recommendations from billions of items within milliseconds~\cite{ma2020off}. 
The first stage focuses on efficiently generating a small set of candidates that contains enough relevant items. To achieve the necessary efficiency, models used in the first stage may be less accurate and biased. 
The second stage only considers the candidates selected in the first stage for generating the final recommendations. It can thus be more resource intensive, which allows second-stage models to be more accurate and less biased. 

While much prior work has focused on improving the efficiency and overall effectiveness of two-stage pipelines~\cite{covington2016deep,ma2020off,kang2019candidate,hron2021component}, less attention has been given to the fairness aspects of this two-stage architecture. We therefore investigate methods for ensuring fair allocation of exposure to the items and their providers in two-stage pipelines, for which there are ample ethical~\cite{kay2015unequal}, economical (\eg, provider retention, super-star economics~\cite{mehrotra2018towards}), and legal (\eg, anti-trust law~\cite{scott2017google}) reasons. 
We specifically investigate how the first stage impacts the fairness of the recommendations, making our work complementary to the existing body of work on fairness and diversity of the second stage in bandits~\cite{joseph2016fairness,chen2020fair,patil2020achieving,gillen2018online,wang2021fairness,schumann2022group} and rankings ~\cite{radlinski2008learning,KuleszaT2012,zehlike2017fa,CelisSV2018ranking,singh2018fairness,biega2018equity,geyik2019fairness,mehrotra2018towards,wang2021user,diaz2020evaluating,wu2021tfrom,do2021two,pitoura2021fairness,kletti2022introducing,Jeunen2021topk,beutel2019fairness,heuss2022fairness}.  
These second-stage methods do not apply to the first stage, since their computation overhead is at least linear in the number of items, which would lead to unacceptable latency in the first stage\footnote{The first-stage recommenders typically employ approximate algorithms to retrieve approximately top-scored items with sub-linear (in the number of items) time complexity, \eg, locality-sensitive hashing~\cite{covington2016deep,ying2018graph,indyk1998approximate}. }. 

We consider a group-based notion of fairness to the items. 
Since the second-stage recommender makes the final recommendations from the candidate set produced by the first stage, a key requirement for the first stage is to select enough relevant items from each group of items to avoid generating an irrecoverably unfair set of candidates. 
For example, consider an e-commerce recommender system, where we aim to ensure both small businesses and large businesses receive an equitable amount of exposure to the users. 
Without a fairness-aware candidate generation policy in the first stage, it might happen that the group of items belonging to small businesses are disproportionately selected less in the first stage, which might be due to biased relevance estimation towards the items from small businesses. In this case, there is little hope for the second-stage recommendation policy to ensure fairness, since (1) there might not be enough relevant items from small businesses for the second-stage recommendations to be fair; (2) second-stage recommendation policies typically only ensure fairness proportional to the items selected in the first stage, where small businesses are already unfairly represented. These fairness issues can appear in almost any two-stage recommender system where we need to consider fair allocation of exposure to the items, including those in hiring, online streaming, and social media. 

In this paper, we study how to ensure fairness with distribution-free and finite-sample guarantees in the first stage of two-stage recommender systems, while retaining the efficiency of existing first-stage recommender systems. In particular, limited by the latency requirements and motivated by the Rooney rule~\cite{collins2007tackling}, we focus on constructing threshold-based first-stage candidate generation policies that can provably select the smallest sets of candidates that contain a desired expected number of relevant items from each group, given any---possibly biased---relevance model.
These guarantees make our approach different from existing works that rely on reducing the bias of relevance estimation in recommendation policies to improve fairness~\cite{burke2018balanced,rastegarpanah2019fighting,yao2017beyond}. While these are certainly useful for the first stage, they do not provide finite-sample and distribution-free guarantees on the fairness and quality of the items selected in the first stage. 

\textbf{Contributions.} We formalize fairness objectives for the first stage of recommender pipelines and propose two threshold selection rules---which we call the union rule and the monotone rule---motivated by distribution-free uncertainty quantification methods. We show that, given any relevance model of queries and items, they can provably select a desired number of relevant items in expectation from each group with high probability, while minimizing the candidate-set size. This result holds even if the relevance model is biased against some groups. We also provide both finite-sample and asymptotic analysis on how close the thresholds selected by the two rules are to the optimal ones. The threshold selection rules and the near-optimality analysis rely on lower and upper confidence bounds on the expected number of relevant items from each group for candidate generation policies. Thus, we derive such confidence bounds from potentially biased and partial user feedback data (\eg, user clicks), which are abundant in many recommender systems. From these bounds, we show that the two threshold selection rules approach the optimal thresholds asymptotically. In addition, we also discuss how the proposed first-stage recommendation policy design can shift the cost of inaccurate relevance estimation and lack of data from the disadvantaged groups\footnote{Disadvantaged groups in this paper refer to groups of items for which the relevance estimation is inaccurate/biased, and/or we lack user feedback data. } to the latency of the second-stage recommender, which provides economic incentives for the decision makers to construct more accurate relevance models and to collect more data for every group of items. 

Finally, we corroborate the theoretical analysis of the two proposed selection rules with an empirical evaluation on the Microsoft Learning to Rank dataset~\cite{QinL13} against several baselines. The results show that only the two proposed selection rules can consistently select enough relevant items from each group across different amounts of user feedback data and accuracies of the relevance model. With a decent amount of data, the two proposed selection rules achieve the smallest candidate-set size among the methods that can select enough relevant items. We also conduct ablation studies to test their robustness to the parameters in the selection rules. 
The code for the empirical evaluation is accessible at \url{https://github.com/LequnWang/Fair-Two-Stage-Recommender}. 


\section{Further Related Work}

Our proposed threshold selection rules are inspired by distribution-free uncertainty quantification methods~\cite{gupta2020distribution}, including calibration~\cite{dawid1982well,platt1999probabilistic} and conformal prediction~\cite{vovk2005algorithmic}. The goal of distribution-free uncertainty quantification is to provide point estimates with finite-sample distribution-free error guarantees (calibration) or confidence intervals (conformal prediction) of some target parameters of interest. In this context, the most relevant work is arguably by Bates \ea~\cite{bates2021distribution}, where they provide a strategy to control the risk of prediction sets from a pool of candidates. Our proposed monotone threshold selection rule uses ideas similar to their strategy. The strategy relies on a point-wise lower confidence bound on the risk, which they derive from full-information data. In contrast, we derive the confidence bounds using partial and biased user feedback, which we can typically have easy access to in recommender systems. In addition, we also provide both finite-sample and asymptotic near-optimality analyses of the two proposed threshold selection rules. 

The confidence bounds we derive are built upon literature on off-policy evaluation in recommender systems~\cite{bottou2013counterfactual,swaminathan2015batch,schnabel2016recommendations,joachims2021recommendations}. Many works have proposed estimators to estimate the expected utility of a contextual bandit~\cite{Dudik2011doubly,su2019cab,wang2017optimal,kallus2018balanced,swaminathan2015self} or a ranking policy~\cite{joachims2017unbiased,yang2018unbiased,wang2016learning,su2019cab,oosterhuis2022doubly} from biased and partial user feedback. Some works have derived finite-sample confidence bounds on the estimation error in contextual bandits~\cite{bottou2013counterfactual,kuzborskij2021confident}. We use similar techniques to derive confidence bounds around the clipped inverse propensity weighted estimator~\cite{ionides2008truncated,bottou2013counterfactual} for the ranking setting. 

Threshold selection rules have also been applied to screening processes~\cite{corbett2017algorithmic,sahoo2021reliable, wang2022improving}. However, these works assume that the candidates are independent and identically distributed (\iid). In contrast, we consider recommendation scenarios where the relevances of the items are dependent given a recommendation request. Thus these approaches do not apply here. 

\section{Fairness in the First Stage}

We consider recommendation problems with $n$ items\footnote{We use $\sbr{\cdot}$ to denote the set $\cbr{1, 2, \ldots, \cdot}$. }  $\db = \rbr{d^j}_{j\in[n]}$, where each item $d^j$ belongs to the item space $\Dcal$, \ie, $d^j\in\Dcal$ for all $j\in[n]$. 
We want to select enough relevant items from several (possibly overlapping) groups $\Gcal$ of items. 
We assume the group information is known for every item $d^j$ and can be included in the feature representation of the item. 
We model the distribution of requests coming into the recommender system using a distribution $P_{Q, \Rb}$ over queries and relevance vectors. 
For each recommendation request, the query $q\in\Qcal$ and the relevance vector $\rb = \rbr{r^j}_{j \in [n]}$ of the items are independently drawn from $q, \rb \sim P_{Q,\Rb}$, where $r^j \in \cbr{0, 1}$ is the relevance of item $d^j$ to the recommendation request. 
The first-stage recommender relies on an expected-relevance estimation model\footnote{We call it the relevance model interchangeably. } $f : \Qcal\times\Dcal\rightarrow [0,1]$, which maps a query $q$ and an item $d^j$ to an estimate $f\rbr{q, d^j}$ of the expected relevance\footnote{We use capital letters to denote random variables and lower case letters to denote realizations of random variables. } $\EE\sbr{R^j \given Q=q}$. 
Given a fixed first-stage relevance model $f$, a first-stage candidate generation policy $\pi^f: \Qcal \times \Dcal^n \rightarrow \{0, 1\}^n$ maps a query $q$ and the items $\rbr{d^j}_{j \in [n]}$ to the selection decisions $\sbb = \pi\rbr{q, \db}$, where $\sbb = \rbr{s^j}_{j \in [n]}$ and
$s^j\in\cbr{0, 1}$ represents whether the item $d^j$ is selected ($s^j = 1$) or not selected ($s^j = 0$). 

Ideally, we would like a policy $\pi^f$ that selects only items that are relevant to the recommendation request from each group. 
Unfortunately, as long as there is no deterministic mapping between the query $q$ and the relevance vector $\rb$, such a perfect candidate generation policy does not exist in general. 
What is worse, to satisfy the latency requirements, recommender systems require that $f$ be computed efficiently, often at the expense of accuracy and unbiasedness. 
So instead, we focus on constructing a policy $\pi^f$ that creates sets of items that are \emph{near-optimal} in terms of the candidate-set size, while provably containing \emph{enough} relevant items in expectation for each group of items, without making assumptions on the distribution $P_{Q, \Rb}$ nor the accuracy/bias of the relevance model $f$. 

In particular, given any relevance model $f$, we consider group-aware threshold policies that select $t_g \in [t_g^{\textnormal{max}}]$ top-scored (predicted by the relevance model $f$) items from each group $g$. $t_g^{\textnormal{max}}$ is the largest candidate-set size the decision makers consider for a group $g$, which conveys the decision makers' belief that there must be enough relevant items from group $g$, had they selected $t_g^{\textnormal{max}}$ top-scored items from $g$. Note that considering only thresholds up to $t_g^{\textnormal{max}}$ provides both computational and statistical efficiency as discussed later. Formally, let $\tb = \rbr{t_g}_{g \in \Gcal}$, a threshold policy $\pi^f_{\tb}$ selects an item if it is among the $t_g$ top-scored items from a group $g$, \ie, 
\begin{equation}
\label{eq:decision_rule_group_threshold_policy}
s^j =
\begin{cases}
1 & \textnormal{if} \, \exists g \in \Gcal \textnormal{ \emph{s.t.} } j \in \cbr{\sigma_{q,g}^f(j'): j' \in \sbr{t_g}} \\
0 & \textnormal{otherwise},
\end{cases}
\end{equation}
where $\sigma_{q, g}^f$ is a ranking of the items in the group $g$ by their estimated relevance to the query $q$ predicted by $f$, which returns the index $\sigma_{q, g}^f(j)$ of the item that is ranked at position $j$ from group $g$ . 
We assume that $\sigma_{q,g}^f$ is deterministic without loss of generality.

We aim to select a threshold vector $\tb\in \prod_{g \in \Gcal}\sbr{t_g^{\textnormal{max}}}$ such that the expected number of relevant items from each group $g \in \Gcal$ is greater than a target $U^\star_g \in \RR$ specified by the decision makers, \ie, 
\begin{equation}
\label{eq:expected_number_relevant_items}
    U_g\rbr{t_g} \coloneqq U_g\rbr{\pi^f_{\tb}} =  \EE_{q, \rb \sim P_{Q, \Rb}} \sbr{\sum_{j \in \sbr{t_g}} r^{ \sigma^f_{q,g}(j)}} \geq U^\star_g, 
\end{equation}
while minimizing the candidate-set size $t_g$ of each group $g$. Note that when the groups are disjoint, minimizing the candidate-set size of each group is equivalent to minimizing the candidate-set size of the whole first-stage recommender. Throughout the paper, we make the following mild assumption on $t_g^{\textnormal{max}}$.
\begin{assumption}
\label{aspt:t-max-enough}
For any group $g\in\Gcal$, 
$U_g\rbr{t_g^{\textnormal{max}}} \geq U_g^\star>0 $. 
\end{assumption}

The target expected numbers of relevant items
$\rbr{U_g^\star}_{g \in \Gcal}$ reflect the decision makers' belief that they can build a fair second-stage recommender system given $U_g^\star$ relevant items from each group $g$. 


\section{Fair and Near-Optimal Threshold Selection Rules}
In this section, we first introduce two threshold selection rules that can provably select enough relevant items in expectation from each group with high probability, while achieving near-optimal candidate-set sizes.  
Both rules rely on a point-wise lower confidence bound on the expected number of relevant items for each threshold policy and each group, and their near-optimality gaps further depend on a point-wise upper confidence bound on that quantity. 
Thus, we derive such lower and upper confidence bounds from some user feedback data (\eg, user clicks), which might be partial and biased. From these bounds, we instantiate concrete threshold selection algorithms, provide asymptotic analysis on how close the proposed threshold selection rules are to the optimal, and discuss how these two rules can incentivize the decision makers to improve the accuracy of the relevance model and collect more data for the disadvantaged groups. 

\subsection{Threshold Selection Rules}
\label{sec:threshold_selection}
For now, lets assume that we have access to a point-wise lower confidence bound $\hat{U}^-_g(t, \alpha)$ on the expected number of relevant items $U_g(t)$ such that for any group $g \in \Gcal$, threshold\footnote{We use $[0:\cdot]$ to denote the set $\cbr{0,1,\ldots,\cdot}$. } $t \in \sbr{0: t_g^{\textnormal{max}} - 1}$, and $\alpha \in (0, 1)$, 
\begin{equation*}
    \Pr\rbr{U_g(t) \geq \hat{U}^-_g(t, \alpha)} \geq 1 -\alpha.
\end{equation*}
We will derive one such bound using user feedback data in Section~\ref{sec:confidence_bound}. 
Given this bound, we propose two threshold selection rules which can ensure that the expected number of relevant items is above the target level $U^\star_g$ with high probability, while achieving near-optimal expected candidate-set size for a group $g$. 

The first rule we propose is called the {\em union threshold selection rule}. Given any success probability $1 - \alpha \in (0, 1)$, it selects the smallest threshold $\hat{t}_g^{\textnormal{union}}$ for a group $g$ such that the lower confidence bound on the expected number of relevant items with failure probability $\frac{\alpha}{t_g^{\textnormal{max}} - 1}$ is greater than the target, \ie,
\begin{equation}
\label{eq:threhold_selection_union}
\hat{t}^{\textnormal{union}}_g \coloneqq \min \cbr{t \in \sbr{t_g^{\textnormal{max}} - 1} : \hat{U}_g^-\rbr{t, \frac{\alpha} {\rbr{t_g^{\textnormal{max}}-1}}} \geq U_g^\star},
\end{equation}
where we define the minimum over the empty set for a group $g$ to be $t_g^{\textnormal{max}}$, \ie, we set the threshold to be $t_g^{\textnormal{max}}$ if none of the lower bounds exceeds the target. 
We show that the union threshold selection rule can select enough relevant items for a group $g$ with high probability in the following theorem, with a proof that applies the union bound over the lower confidence bounds for each threshold, as reflected in our naming of the rule. 

\begin{theorem}
\label{theo:threshold_selection_union}
Under Assumption~\ref{aspt:t-max-enough}, for any group $g \in \Gcal$ and $\alpha \in (0, 1)$, with probability at least $1 - \alpha$,
\[
U_g\rbr{\hat{t}_g^{\textnormal{union}}} \geq U_g^\star. 
\]
\end{theorem}
\begin{proof}
For a group $g$, applying the union bound to the lower confidence bounds for each threshold, and by Assumption~\ref{aspt:t-max-enough}, we have that with probability at least $1 - \alpha$, 
\[
U_g(t) \geq \hat{U}^-_g\rbr{t, \frac{\alpha}{\rbr{t_g^\textnormal{max} - 1}} } \quad \forall t \in \sbr{t_g^{\textnormal{max}}}.  
\]
When the above event holds, 
\[
U_g\rbr{\hat{t}_g^{\textnormal{union}}} \geq \hat{U}^-_g\rbr{\hat{t}_g^{\textnormal{union}}, \frac{\alpha}{\rbr{t_g^\textnormal{max} - 1}}} \geq U_g^\star, 
\]
where the second inequality is by the definition of $\hat{t}_g^{\textnormal{union}}$. 
\end{proof}

The second selection rule is called the {\em monotone threshold selection rule}. Given any success probability $1 - \alpha \in (0,1)$, it selects the smallest threshold $\hat{t}_g^{\textnormal{mono}}$ for a group $g$ such that the lower confidence bound with failure probability $\alpha$ for every threshold larger than or equal to $\hat{t}_g^{\textnormal{mono}}$ is greater than the target $U_g^\star$, \ie, 
\begin{equation}
\label{eq:threshold_selection_monotone}
    \hat{t}_g^{\textnormal{mono}} \!\coloneqq\! \min\! \cbr{t \!\in\! \sbr{t_g^{\textnormal{max}} - 1}\! :\! \hat{U}^-_g\!\rbr{t', \alpha} \! \geq\! U_g^\star, \forall t \leq t' \! < \!t_g^{\textnormal{max}}}.  
\end{equation}

Compared to $\hat{t}_g^\textnormal{union}$, $\hat{t}_g^\textnormal{mono}$ allows for a larger failure probability in the confidence lower bounds for each threshold, but requires that the lower bounds be greater than the target for all the thresholds larger than the selected one, in addition to the selected one. It can also ensure selecting enough relevant items with high probability for any group as shown in the following theorem, where the proof leverages the fact that $U_g$ is monotonically increasing, as reflected in our naming of the rule. 

\begin{theorem}
\label{theo:threshold_selection_monotone}
Under Assumption~\ref{aspt:t-max-enough}, for any group $g \in \Gcal$ and $\alpha \in (0, 1)$, with probability at least $1 - \alpha$, 
\[
U_g\rbr{\hat{t}_g^{\textnormal{mono}}} \geq U_g^\star. 
\]
\end{theorem}

\begin{proof}
The proof of the theorem is inspired by that of Theorem $1$ in~\cite{bates2021distribution}. 

For any group $g$, let $t_g^\star$ be the smallest threshold such that the expected number of relevant items from group $g$ using the threshold policy is larger than or equal to $U_g^\star$, \ie,
\begin{equation}
\label{eq:optimal_threshold}
t_g^\star \coloneqq \min \cbr{t \in \sbr{t_g^{\textnormal{max}}}: U_g(t) \geq U_g^\star}. 
\end{equation}
By Assumption~\ref{aspt:t-max-enough}, we know that the set on the right of Eq.\ref{eq:optimal_threshold} is non-empty. Suppose $U_g\rbr{\hat{t}_g^{\textnormal{mono}}} < U_g^\star$, we know $U_g\rbr{\hat{t}_g^{\textnormal{mono}}} < U_g^\star \leq U_g\rbr{t^\star_g}$ by the definition of $t^\star_g$. Thus $\hat{t}_g^{\textnormal{mono}} < t_g^\star$ by the fact that $U_g$ is monotonically increasing. Since $\hat{t}_g^{\textnormal{mono}}$ and $t_g^\star$ are integers, we have $t_g^\star - 1 \geq  \hat{t}_g^{\textnormal{mono}}$. By the definitions of $\hat{t}_g^{\textnormal{mono}}$ and $t_g^\star$, this further implies that $t_g^\star > 1$ and
\[
\hat{U}_g^-\rbr{t^\star_g - 1, \alpha} \geq U^\star_g > U_g\rbr{t_g^\star - 1}. 
\]
By the lower confidence bound, we know that this happens with probability at most $\alpha$, which concludes the proof.  
\end{proof}

We summarize the fair first-stage threshold-policy selection algorithm in Algorithm~\ref{alg:fair_first_stage_policy}. 
\begin{algorithm}[t]
\begin{algorithmic}[1]
\STATE{{\bf input:} $\Gcal$, $\rbr{U_g^\star}_{g \in \Gcal}$, $\rbr{t_g^{\textnormal{max}}}_{g \in \Gcal}$, $\rbr{\hat{U}_g^{-}}_{g \in \Gcal}$, $\alpha$}
\STATE{Compute the lower confidence bounds for each group $g \in \Gcal$ and each threshold $t \in \sbr{t_g^{\textnormal{max}} - 1}$: $\hat{U}^-_g\rbr{t, \alpha}$ for $\hat{t}_{g}^{\textnormal{mono}}$ or $\hat{U}^-_g\rbr{t, \frac{\alpha}{t_g^{\textnormal{max}} - 1}}$ for $\hat{t}_{g}^{\textnormal{union}}$. }
\STATE{Compute $\hat{t}_g$ for each group $g \in \Gcal$ : $\hat{t}_g = \hat{t}_g^{\textnormal{mono}}$ by Eq~\ref{eq:threshold_selection_monotone} or $\hat{t}_g = \hat{t}_g^{\textnormal{union}}$ by Eq~\ref{eq:threhold_selection_union}. }
\STATE{{\bf return} $\pi^f_{\hat{\tb}}$, where $\hat{\tb} = \rbr{\hat{t}_g}_{g \in \Gcal}$. }
\end{algorithmic}
\caption{Fair First-Stage Threshold-Policy Selection}
\label{alg:fair_first_stage_policy}
\end{algorithm}

\subsection{Finite-Sample Near-Optimality Gaps}
So far, we have shown that both threshold selection rules ensure that the expected number of relevant items is large enough with high probability from each group. Now we characterize how far the expected number of relevant items selected by each threshold policy deviates from the target $U^\star_g$ for each group $g$. 

The finite-sample near-optimality gaps we prove depend also on a point-wise upper confidence bound $\hat{U}^+_g\rbr{t, \alpha}$ on the expected number of relevant items $U_g(t)$ such that for any group $g \in \Gcal$, threshold $t \in \sbr{t_g^{\textnormal{max}} - 1}$, and $\alpha \in (0, 1)$, 
 \begin{equation}
     \Pr\rbr{U_g(t) \leq \hat{U}_g^+\rbr{t, \alpha}} 
 \geq 1 - \alpha. 
 \end{equation}
We will show how to derive one such bound for data that takes the form of partial-information user feedback in Section~\ref{sec:confidence_bound}. 

For the union threshold selection rule $\hat{t}_g^{\textnormal{union}}$, we have the following proposition that bounds how much more relevant items it selects than the target $U_g^\star$. 

\begin{proposition}
\label{prop:optimality_gap_union}
Under Assumption~\ref{aspt:t-max-enough}, for any group $g \in \Gcal$ and $\alpha \in (0, 1)$, with probability at least $1 - \alpha$,
\begin{multline*} U_g\rbr{\hat{t}_g^{\textnormal{union}}} - U_g^\star  < \hat{U}^+_g\rbr{\hat{t}_g^{\textnormal{union}},\frac{\alpha}{t_g^{\textnormal{max}} - 1}} \\ - \hat{U}_g^-\rbr{\hat{t}_g^{\textnormal{union}} -1, \frac{\alpha}{t_g^{\textnormal{max}} -1}}. 
\end{multline*}
\end{proposition}

The complete proofs of this and the following propositions are on the arxiv\footnote{\url{https://arxiv.org/abs/2205.15436}}. We can similarly derive the finite-sample near-optimality gap for $\hat{t}_g^{\textnormal{mono}}$ as shown in the following proposition. 

\begin{proposition}
\label{prop:optimality_gap_monotone}
Under Assumption~\ref{aspt:t-max-enough}, for any group $g$ and $\alpha \in (0, 1)$, with probability at least $1 - \alpha$, 
\begin{multline*} U\rbr{\hat{t}_g^{\textnormal{mono}}} - U_g^\star  < \hat{U}^+_g\rbr{\hat{t}_g^{\textnormal{mono}}, \frac{\alpha}{t_g^{\textnormal{max}} - 1}} - \hat{U}_g^-\rbr{\hat{t}_g^{\textnormal{mono}} -1, \alpha}. 
\end{multline*}
\end{proposition}

The proof of proposition~\ref{prop:optimality_gap_monotone} is almost the same as that of Proposition~\ref{prop:optimality_gap_union}, and therefore we omit it. 
We can see that both threshold selection rules rely on the lower confidence bounds, and their optimality gaps depend further on the upper confidence bounds on the expected number of relevant items for threshold policies. To instantiate the two rules and their finite-sample near-optimality gaps,  
we introduce how to derive such bounds using user feedback data in the next section.

\subsection{Confidence Bounds from User Feedback}
\label{sec:confidence_bound}

In many recommender systems, we have access to an abundance of user feedback (\eg, user clicks, dwell times) that can help reveal the relevance of the items to the queries. 
However, these data are partial (we only observe a user's feedback for a subset of the items) and biased (by the presentation of the policy that logged the data). Thus, we derive confidence bounds that are robust to these properties\footnote{The full-information setting is a special case of this partial-information setting where the users observe every item. }. 

More formally, we use a sample of logged user feedback for $m$ recommendations served by a deployed logging policy $\pi_0$. 
For each recommendation request $i$, we merely assume that the query and the relevance vector are independently sampled from the query and relevance-vector distribution $q_i, \rb_i \sim P_{Q, \Rb}$. 
Instead of directly observing the relevance vector $\rb_i$, we observe some user feedback (\eg, user clicks) $\cbb_i = \rbr{c^j_i}_{j \in [n]}$ that is typically biased by the recommendation that $\pi_0$ made (e.g., position in ranking). 
To model this presentation bias, we follow the standard approach \cite{joachims2017unbiased, agarwal2019estimating, fang2019intervention,ai2018unbiased,wang2018position} where the user feedback for an item $d^j$ is decomposed as the product of the user observation and the relevance $c_i^j = o_i^j r_i^j$, where $o_i^j \in \cbr{0, 1}$ denotes whether the user observes the item. 
The user observation vector $\ob \in \cbr{0, 1}^n$ is generated from $P^{\pi_0}_{\Ob \given Q, \Rb}$. 
We call the conditional probability that the user observes an item under the logging policy $\pi_0$ the {\em propensity}, and denote it as $p^j_i \coloneqq \Pr\rbr{O_i^j = 1 \given Q = q_i, \Rb = \rb_i; \pi_0}$. 
In the case of one-item recommendation, this can be interpreted as the probability that the logging policy $\pi_0$ recommends the item. Since we control the logging policy, this propensity is known by design. 
In the case of ranking, the propensities typically need to be estimated. There are many existing works on estimating the propensities~\cite{joachims2017unbiased, agarwal2019estimating, fang2019intervention,ai2018unbiased,wang2018position} by making assumptions on how the users interact with the ranked items (\eg, position-based click models~\cite{agarwal2019estimating} where the propensity only depends on the rank of the item, and cascade click models~\cite{chandar2018estimating,vardasbi2020cascade} where it further depends on the relevances of the items in a particular way).  
Thus we assume we know the (estimated) propensities, and the batch of logged user feedback for constructing the confidence bounds is $\Scal_{\textnormal{CB}} = \{q_i, \cbb_i, \pb_i\}_{i\in[m]}$, where $\pb_i = \rbr{p_i^j}_{j \in [n]}$ and $m > 1$. Throughout the paper, we assume that the propensities are positive, as formally described below, which can be achieved by carefully designing the logging policy. 
\begin{assumption}
\label{aspt:positive-propensity}
There exists $\gamma > 0$ such that
\[
p^{\sigma_{q,g}^f(j)} > \gamma \quad \forall g \in \Gcal, q \in \Qcal, j \in \sbr{t_g^{\textnormal{max}}}. 
\]
\end{assumption}

The confidence bounds we derive using $\Scal_{\textnormal{CB}}$ are based on the clipped inverse propensity weighted (CIPW) estimator~\cite{horvitz1952generalization,ionides2008truncated,bottou2013counterfactual} adapted to this ranking setting. The following CIPW estimator estimates the expected number of relevant items $U_g(t_g)$ of threshold policies (recall that we use capital letters to denote random variables)
\begin{equation}
\label{eq:CIPW_estimator}
\hat{U}^{\textnormal{CIPW}}_{g, \lambda}\rbr{t_g} \coloneqq \frac{1}{m} \sum_{i\in[m]} \sum_{j\in\sbr{t_g}} \min\rbr{\lambda, \frac{1}{p^{\sigma_{Q_i, g}^f(j)}_i}}C^{\sigma_{Q_i, g}^f(j)}_i, 
\end{equation}
with a clipping parameter $\lambda$---the maximum inverse propensity weight---to balance the bias and variance of the estimator. 
The following propositions provide empirical upper and lower confidence bounds on the expected number of relevant items $U_g\rbr{t_g}$ around its CIPW estimate $\hat{U}^{\textnormal{CIPW}}_{g, \lambda}\rbr{t_g}$. 

\begin{proposition}
\label{prop:lower_bound}
Under Assumption~\ref{aspt:positive-propensity}, for any $\lambda > 0$, $g \in \Gcal$, $t \in \sbr{t_g^{\textnormal{max}} - 1}$, and $\alpha \in (0, 1)$, with probability at least $1 - \alpha$, we have that
\begin{multline}
\label{eq:lower_confidence_bound}
U_g\rbr{t} \geq \hat{U}^{\textnormal{CIPW}}_{g, \lambda}\rbr{t} -  \sqrt{\frac{2V_m\rbr{\Zb^{g, t} } \ln(2/\alpha)}{m}}   - \frac{7t\lambda\ln(2/\alpha)}{3(m - 1)} \\ \coloneqq \hat{U}_g^-\rbr{t, \alpha}, 
\end{multline}
where $\Zb^{g, t} = \rbr{Z^{g, t}_i}_{i \in [m]}$ with
\[
Z_i^{g, t} = \sum_{j\in[t]} \min\rbr{\lambda, \frac{1}{p^{\sigma_{Q_i, g}^f(j)}_i}}C^{\sigma_{Q_i, g}^f(j)}_i
\]
be the CIPW estimate of recommendation request $i$, and
\[
V_m\rbr{\Zb^{g, t}} = \frac{1}{m(m - 1)}\sum_{1 \leq i < j \leq m}\rbr{Z_i^{g, t} - Z_j^{g, t}}^2 
\]
be the sample variance. 
\end{proposition}

\begin{proposition}
\label{prop:upper_bound}
Under Assumption~\ref{aspt:positive-propensity}, for any $\lambda > 0$, $g \in \Gcal$, $t \in \sbr{t_g^{\textnormal{max}} - 1}$, and $\alpha \in (0, 1)$, with probability at least $1 - \alpha$, we have that 
\begin{multline*}
\label{eq:upper_confidence_bound}
U_g(t) \leq \hat{U}^{\textnormal{CIPW}}_{g, \lambda}(t) + \sqrt{\frac{2V_m\rbr{ \Zb^{g, t}} \ln(4/\alpha)}{m}} + \frac{7t\lambda \ln(4/\alpha)}{3(m - 1)} \\ + t\sqrt{\frac{\ln(2/\alpha)}{2m}} +  \frac{1}{m} \sum_{i \in [m]}\sum_{j \in [t]}\max\rbr{0, 1 - \lambda p_i^{\sigma_{Q_i, g}^f(j)}} \coloneqq \hat{U}^+(t, \alpha). 
\end{multline*}

\end{proposition}


Note that both bounds apply to any group of items including the group of all items $\db$. Though we apply them for fair first-stage threshold-policy design and characterizing the finite-sample near-optimality gaps, we believe that they might be of independent interests in other contexts. 

With these bounds, we can instantiate Algorithm~\ref{alg:fair_first_stage_policy} and the finite-sample near-optimality gaps. We can see that the two selection rules can ensure selecting enough relevant items from each group regardless of the accuracy of the relevance model nor the amount of user feedback data across groups. However, the candidate-set sizes of the groups are determined by those factors. In particular, we need to include more items from a group if the relevance model is less accurate and/or we have less data for the group. This shifts the costs of inaccurate relevance estimation and/or lack of data for the items from disadvantaged groups to the latency of the second-stage recommender, and thus provides economic incentives for the decision makers to build accurate relevance models and collect enough data for every group. 

\subsection{Asymptotic Near-Optimality Analysis}
From these upper and lower confidence bounds, we can now analyze how close the selected thresholds are to the optimal thresholds $t_g^\star$ as defined in Eq.~\ref{eq:optimal_threshold} asymptotically. In the following propositions, we show that the expected number of relevant items using the selected policies will converge to the target asymptotically, and the selected thresholds will also converge to the optimal thresholds asymptotically, under mild assumptions. 

\begin{proposition}
\label{prop:near_optimality_num_relevant}
Let $\hat{t}_g$ be either $\hat{t}_g^{\textnormal{union}}$ or $\hat{t}_g^{\textnormal{mono}}$, and $\lambda = \sqrt{m}$. Under Assumption~\ref{aspt:t-max-enough} and ~\ref{aspt:positive-propensity}, for any group $g \in \Gcal$ and any $\alpha \in (0, 1)$, it holds almost surely (with probability $1$) that for any $\delta > 0$, there exists $c > 0$ such that for any $m > c$, 
\[
U^\star_g - \delta <  U_g\rbr{\hat{t}_g} < 1 + U_g^\star + \delta,
\]
and even stronger than the second inequality above, 
\[
U_g\rbr{\hat{t}_g - 1} < U_g^\star + \delta. 
\]
\end{proposition}

\begin{proposition}
\label{prop:near_optimality_threshold}
Under the conditions in Proposition~\ref{prop:near_optimality_num_relevant} and further assume that $U_g$ is strictly increasing,
we have that for any group $g \in \Gcal$, $\alpha \in (0, 1)$, it holds almost surely that there exists $c > 0$ such that for any $m > c$, 
\[
t_g^\star \leq \hat{t}_g \leq t_g^\star + 1. 
\]
\end{proposition}
%

\section{Empirical Evaluation}

In this section, we compare the union $\hat{t}_g^{\textnormal{union}}$ and the monotone $\hat{t}_g^{\textnormal{mono}}$ threshold selection ruls with several competitive baselines on first-stage recommendation scenarios simulated from the Microsoft Learning-to-Rank WEB30K dataset~\cite{QinL13}. 
\subsection{Experiment Setup}
The dataset consists of $30,000$ queries, along with the relevances and the features of the items per query. We divide the items into two categories---``old and estabilished'' and ``new and undiscovered''---by their ``url click count'' feature given in the dataset. The feature represents ``the click count of a url aggregated from user browsing data in a period''. We set the items with zero click count as the disadvantaged group \emph{disadv} and the other items as the advantaged group \emph{adv}. We binarize the relevance by assigning relevance $1$ to items with an original label of 2, 3, or 4, and $0$ to the others. The average numbers of relevant items per query from each group are $\textnormal{AR}_{\textnormal{adv}} = 6.16$ and $\textnormal{AR}_{\textnormal{disadv}} = 13.99$. 

For each experiment, we randomly split the data into $1\%$ for training a logistic regression model as the relevance model $f$,  $69\%$ for simulating the user feedback, and $30\%$ for testing different first-stage candidate generation policies. To simulate user feedback, we follow prior works~\cite{joachims2017unbiased} to assume that users follow a position-based click model~\cite{craswell2008experimental}. More specifically, for each recommendation request $i$, we randomly sample a query from the $69\%$ data, create a ranking for the top $t_g^{\textnormal{max}}$ items from each group by the relevance model $f$, simulate the user observation $o_i^j \sim \textnormal{Bernoulli}\rbr{p_i^j}$ with the propensity set as one over the rank of the item $p_i^j = \frac{1}{rank(j\given \pi_0)}$ for each item $j$, and the simulated user feedback is $c_i^j = o_i^jr_i^j$.  

Unless specified explicitly, we set the size of the user feedback $m = 100,000$, the clipping parameter $\lambda = 100$, and the largest number of items we consider from both groups to be the same $t_{\textnormal{adv}}^{\textnormal{max}} = t_{\textnormal{disadv}}^{\textnormal{max}} = 50$, the success probability $1 - \alpha = 0.9$ by default. We set the target expected numbers of relevant items $U^\star_{\textnormal{adv}}$ and $U^\star_{\textnormal{disadv}}$ to satisfy the equal opportunity constraint~\cite{hardt2016equality,wang2022improving}, \ie, $U^\star_{\textnormal{adv}} / \textnormal{AR}_{\textnormal{adv}} = U^\star_{\textnormal{disadv}} / \textnormal{AR}_{\textnormal{disadv}}$, subject to $ U^\star_{\textnormal{disadv}} + U^\star_{\textnormal{adv}} = 5$. 

\textbf{Baselines. } We compare $\hat{t}_g^{\textnormal{mono}}$ (\emph{CIPW-LB-mono}) and $\hat{t}_g^{\textnormal{union}}$ (\emph{CIPW-LB-union}) with several baselines. The \emph{Uncalibrated Individual} baseline selects top-ranked items from each group until the sum of the scores predicted by $f$ exceed the target. The \emph{Uncalibrated Marginal} baseline selects the smallest threshold for each group using the simulated user data such that the average sum of scores is greater than the target. The \emph{Platt Individual}, \emph{Platt Marginal}, \emph{Platt PG Individual}, and \emph{Platt PG Marginal} baselines are the same as the uncalibrated ones, except that they apply Platt scaling~\cite{platt1999probabilistic} to the relevance model using user feedback data through inverse propensity weighting~\cite{kweon2021obtaining}, where ``PG'' implies that we calibrate the relevance model per group. The \emph{IPW} baseline selects the smallest threshold such that the inverse propensity weighted (IPW) estimator on the expected number of relevant items of the threshold policy is greater than the target for each group. 

\textbf{Metrics. } To compare different first-stage candidate generation policies, we run experiments $50$ times for each setting. For each run, we estimate whether each candidate generation policy selects enough relevant items $\textnormal{ER}_{g} \coloneqq \II \cbr{\hat{\EE}\sbr{\sum_{j \in [n]: d^j\in g} s^j r^j} \geq U_g^\star}$ and the candidate-set size $\textnormal{CSS}_g = \hat{\EE}_g\sbr{\sum_{j \in [n]:d^j\in g} s^j}$ for both $g = \textnormal{adv}$ and $g = \textnormal{disadv}$ on the $30\%$ full-information test data. We then compare different polices in terms of the percentage of times they ensure selecting enough relevant items (along with standard errors) and the average candidate-set size (along with standard deviations) for both groups across the $50$ runs. 
\subsection{How do different methods scale with the size of user feedback data? }

\begin{figure*}[t]
\centering
\includegraphics[width=1.\textwidth]{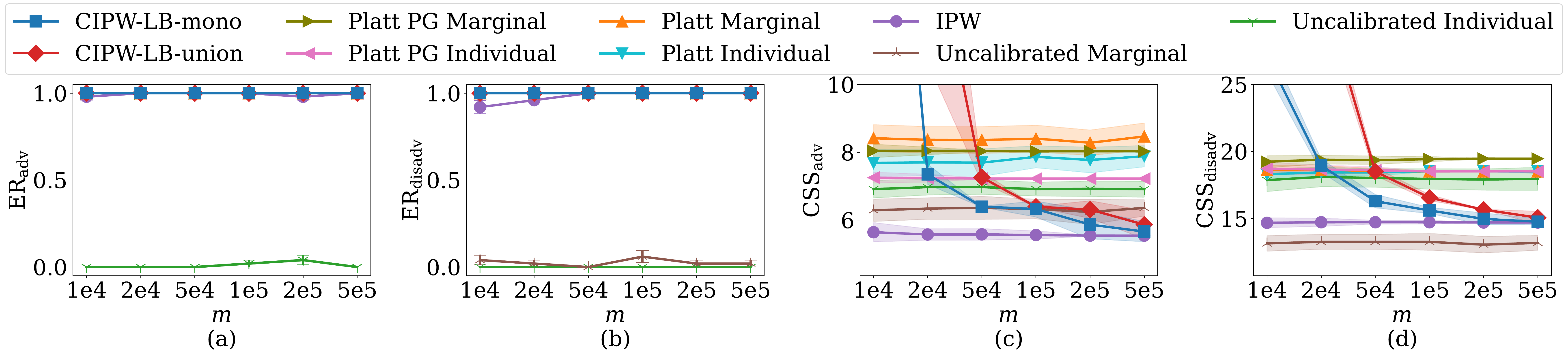}
\vspace*{-4mm}
\caption{Comparison of different first-stage candidate generation policies when we vary the amount $m$ of user feedback data. The left two plots show the empirical probability, along with standard error bars, that each policy selects enough relevant items for the advantaged group $\textnormal{ER}_{\textnormal{adv}}$ and the disadvantaged group $\textnormal{ER}_{\textnormal{disadv}}$ across $50$ runs. The right two plots show the empirical average, along with one standard deviation as shaded regions, of the expected candidate-set size for the advantaged group $\textnormal{CSS}_{\textnormal{adv}}$ and the disadvantaged group $\textnormal{CSS}_{\textnormal{disadv}}$ across 50 runs. 
} 
\label{fig:exp_cal_size}
\end{figure*}

\begin{figure*}[t]
\centering
\includegraphics[width=1.\textwidth]{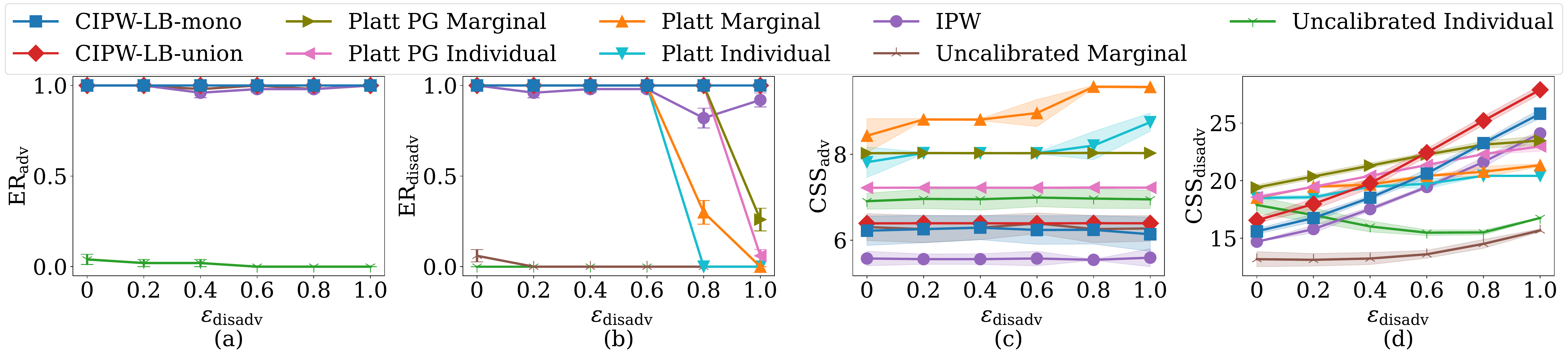}
\vspace*{-4mm}
\caption{Comparison of different first-stage candidate generation policies when we vary the accuracy of the relevance model to the disadvantaged group by changing the relevance noise ratio $\epsilon_{\textnormal{disadv}}$ to the disadvantaged group. 
}
\label{fig:exp_noise_ratio}
\end{figure*}

Figure~\ref{fig:exp_cal_size} compares different candidate generation policies with respect to the percentage of times they selects enough relevant items ((a) and (b)), and the average candidate-set sizes ((c) and (d)) for both the advantaged groups ((a) and (c)) and the disadvantaged group ((b) and (d)). 
We can see that the Uncalibrated baselines do not guarantee selecting enough relevant items in general. The IPW baseline selects enough relevant items for most of the instances, and achieves smaller or comparable candidate-set sizes as the two proposed rules.  Part of the reason is that we consider discrete threshold policies, and the expected number of relevant items of the optimal threshold $U_g\rbr{t_g^\star}$ is typically larger than the target $U_g^\star$. As long as the IPW estimator does not overestimate the expected number of relevant items more than the difference $U_g\rbr{t_g^\star} - U_g^\star$, it will select enough relevant items. However, as we can see, it fails to select enough relevant items more often on the disadvantaged group, especially when the size of the user feedback data is small. This is because the variance of the CIPW estimator is larger for the disadvantaged group due to a larger threshold and smaller propensities. This phenomenon is more obvious when the relevance model is less accurate for the disadvantaged group as shown in the next subsection. Among the methods that always select enough relevant items, the two proposed threshold selection rules have the smallest candidate-set sizes when the amount of data is large. As we have less and less data, the two proposed rules select more and more items to account for the increasing uncertainty due to the lack of data. Comparing the two proposed rules, the monotone selection rule consistently outperforms the union selection rule in terms of the candidate-set size across data sizes, partly because it allows for a larger failure probability in the lower confidence bounds.

\subsection{How does the accuracy of the relevance model affect different groups of items? }

To simulate scenarios where the relevance model $f$ might be less accurate for the disadvantaged group, we vary the accuracy of $f$ for the disadvantaged group by replacing its prediction for items in the disadvantaged group with some noise $\beta$ sampled from $\beta \sim \textnormal{Beta}(1, 10)$, with probability $\epsilon_{\textnormal{disadv}}$, \ie,  $f_{\textnormal{disadv}} = \rbr{1-\eta} f + \eta \beta$, where $\eta \sim \textnormal{Bernoulli}\rbr{\epsilon_{\textnormal{disadv}}}$. 

Figure~\ref{fig:exp_noise_ratio} compares different candidate generation policies when we vary the relevance noise ratio $\epsilon_{\textnormal{disadv}}$ to the disadvantaged group. 
We can see that, as the relevance model becomes less accurate for the disadvantaged group ($\epsilon_{\textnormal{disadv}}$ becomes larger), only the two proposed threshold selection rules always select enough relevant items for the disadvantaged group. In particular, IPW fails to select enough relevant items substantially more often for the disadvantaged group than for the advantaged group, since it does not account for the uncertainty in the estimation process. This highlights the benefits of distribution-free and finite-sample guarantees the proposed union and monotone threshold selection rules enjoy. The two rules achieve this by selecting more items from the disadvantaged group to account for the increasing uncertainty due to the less accurate relevance model for the disadvantaged group. Again, the monotone threshold selection rule consistently outperforms the union threshold selection rule in terms of the candidate-set size across different relevance noise ratios to the disadvantaged group. 

\subsection{How robust are the two proposed rules to the clipping parameter $\lambda$? }

\begin{figure*}[ht]
\centering
\includegraphics[width=1.\textwidth]{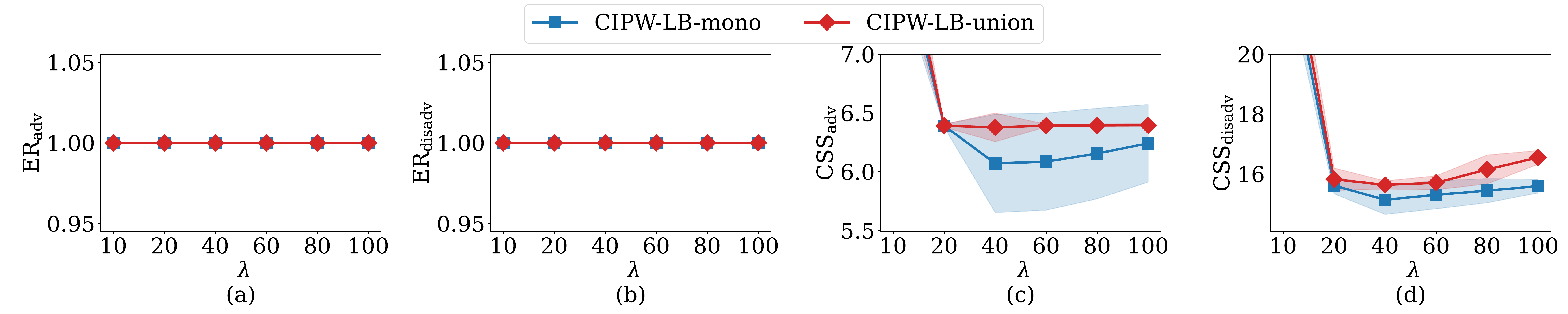}
\vspace{-4mm}
\caption{Analysis of the two proposed threshold selection rules when we vary the clipping parameter $\lambda$. 
}
\label{fig:exp_W_max}
\end{figure*}
\begin{figure*}[ht]
\centering
\includegraphics[width=1.\textwidth]{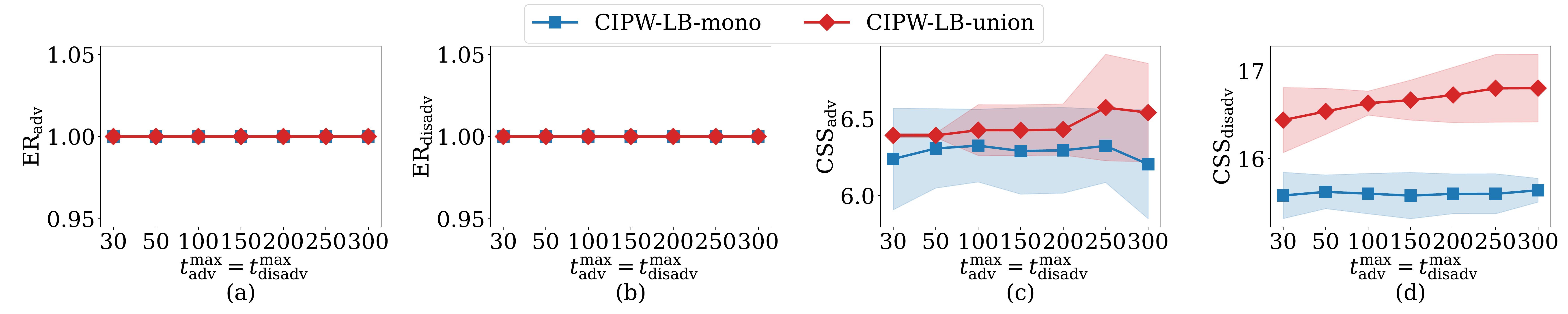}
\vspace{-4mm}
\caption{Analysis of the two proposed threshold selection rules when we vary the largest numbers of items we consider for each group, which we set to be the same $t_{\textnormal{adv}}^{\textnormal{max}} = t_{\textnormal{disadv}}^{\textnormal{max}}$. 
}
\label{fig:exp_t_max}
\end{figure*}

Figure~\ref{fig:exp_W_max} shows the performance of the two proposed threshold selection rules with varying clipping parameter $\lambda$. 
We can see that the two proposed rules can select enough relevant items across different values in the clipping parameter $\lambda$, which confirms their distribution-free and finite-sample guarantees. The candidate-set sizes for both groups exhibit a bowl shape, which is consistent with our theory of finite-sample near-optimality gaps, that the optimal $\lambda$ lies in the middle where there is a favorable bias and variance trade-off in the CIPW estimator. 

\subsection{Are the two proposed rules robust to the largest considered threshold $t_g^{\textnormal{max}}$? }

We compare the performance of the two proposed threshold selection rules when we use different largest considered threshold $t_g^{\textnormal{max}}$ in Figure~\ref{fig:exp_t_max}. Unsurprisingly, the two proposed selection rules still select enough relevant items across different values of $t_g^{\textnormal{max}}$, as predicted by the distribution-free and finite-sample guarantees. In terms of the candidate-set size, there is a slight increase for the union threshold selection rule as the largest considered threshold increases. This is expected since it uses smaller and smaller failure probabilities in the lower confidence bounds. For the monotone threshold selection rule, the candidate-set size does not change with the maximum considered threshold, partly because the failure probability it uses in the lower confidence bounds does not change. This also shows that the lower confidence bounds used in the monotone selection rule are still not too loose even when the thresholds are large, since the threshold $t$ is in the log terms in the bounds.

\section{Conclusion}

In this work, we initiated the study of fairness in the first stage of two-stage recommender systems. In particular, we proposed two threshold-policy selection rules that can select fair first-stage policies using abundantly available user-feedback data even if the relevance model used in the first stage is biased and has disparate accuracy across groups. We show that the two selection rules can provably select enough relevant items in expectation from each group with high probability, achieve near-optimal candidate-set sizes, and retain the efficiency of most existing first-stage recommender systems. Both the theoretical analysis and the empirical evaluation confirm that the two proposed selection rules are robust to the amount of user feedback data, the accuracy of the relevance model, and the parameters inside the two rules. 

\begin{acks}
This research was supported in part by NSF Awards
IIS-1901168 and IIS-2008139. All content represents the
opinion of the authors, which is not necessarily shared or endorsed by their respective employers and/or sponsors.
\end{acks}

\bibliographystyle{ACM-Reference-Format}
\bibliography{fair-first-stage}

\appendix
\onecolumn
\section{Proof of Proposition~\ref{prop:optimality_gap_union}}

\begin{proof}
By applying the union bound over the upper confidence bounds for all the thresholds for a group $g$ and by Assumption~\ref{aspt:t-max-enough}, we have that with probability at least $1 - \alpha$, 
\[
U_g\rbr{t} \leq \hat{U}^{+}_g\rbr{t , \frac{\alpha}{t_g^{\textnormal{max}} - 1}} \quad \forall t \in \sbr{t_g^{\textnormal{max}}}, 
\]
and thus
\begin{multline*}
U_g\rbr{\hat{t}_g^{\textnormal{union}}}  \leq \hat{U}_g^+\rbr{\hat{t}_g^{\textnormal{union}} , \frac{\alpha}{t_g^{\textnormal{max}} - 1}} = \hat{U}_g^-\rbr{\hat{t}_g^{\textnormal{union}} - 1, \frac{\alpha}{t_g^{\textnormal{max}} - 1}}  + \hat{U}_g^+\rbr{\hat{t}_g^{\textnormal{union}}, \frac{\alpha}{t_g^{\textnormal{max}} - 1}}  - \hat{U}_g^-\rbr{\hat{t}_g^{\textnormal{union}} - 1, \frac{\alpha}{t_g^{\textnormal{max}} - 1}}  \\ <  U_g^\star + \hat{U}_g^ + \rbr{\hat{t}_g^{\textnormal{union}}, \frac{\alpha}{t_g^{\textnormal{max}} - 1}} - \hat{U}_g^- \rbr{\hat{t}_g^{\textnormal{union}} - 1, \frac{\alpha}{t_g^{\textnormal{max}} - 1}}, \end{multline*}
where the last inequality is by the definition of $\hat{t}_g^{\textnormal{union}}$ and Assumption~\ref{aspt:t-max-enough}. 
\end{proof} 
\section{Proof of Proposition~\ref{prop:lower_bound}}
\begin{proof}
We can decompose the difference between $U_g(t)$ and its CIPW estimate as the bias and the concentration terms as follows
\begin{equation*}
\label{eq:bias_concentration_decomposition}
U_g(t) - \hat{U}^{\text{CIPW}}_{g, \lambda}(t) = \underbrace{ U_g(t) - U^{\text{CIPW}}_{g, \lambda}(t)}_{\textnormal{Bias}} + \underbrace{ U^{\text{CIPW}}_{g, \lambda}(t) - \hat{U}^{\textnormal{CIPW}}_{g, \lambda}  (t)}_{\textnormal{Concentration}}, 
\end{equation*}
where
\begin{equation*}
U_{g, \lambda}^{\textnormal{CIPW}}(t) \coloneqq \EE_{q, \rb \sim P_{Q, \Rb}, \ob \sim P^{\pi_0}_{\Ob \given q , \rb}}\sbr{z^{g, t}} =\EE_{q, \rb \sim P_{Q, \Rb}}\sbr{\sum_{j \in [t]} \min\rbr{\lambda p^{\sigma_{q, g}^f(j)}, 1}r^{\sigma_{q, g}^f(j)}}. 
\end{equation*}
Lets first look at the concentration term. Since for all $i \in [m]$ and $j \in \sbr{t_g^{\textnormal{max}} - 1}$, $0 \leq \min\rbr{\lambda, \frac{1}{p^{\sigma_{Q_i, g}^f(j)}_i}} \leq \lambda$, and $C^{\sigma_{Q_i, g}^f(j)}_i \in \cbr{0, 1}$, we have
\[
0\leq Z_i^{g, t} \leq t \lambda \quad \forall i\in[m]. 
\]
Applying the empirical Bernstein bound (Theorem 4 in~\cite{MaurerP09}) to the \iid\, random variables $\Zb^{g, t}$, we have that for any $\alpha \in (0, 1)$, with probability at least $1 - \alpha$, 
\[
U^{\text{CIPW}}_{g, \lambda}(t) - \hat{U}^{\textnormal{CIPW}}_{g, \lambda} (t) \geq - \sqrt{\frac{2V_m\rbr{ \Zb^{g, t}} \ln(2/\alpha)}{m}} - \frac{7t\lambda \ln(2/\alpha)}{3(m - 1)}. 
\]

For the bias term, we have
\begin{equation*}
U_g(t) - U^{\text{CIPW}}_{g, \lambda}(t)
= \EE_{q, \rb \sim P_{Q, \Rb}} \sbr{\sum_{j \in [t]} r^{\sigma_{q, g}^f(j)} - \min\rbr{\lambda p^{\sigma^f_{q, g}(j)}, 1}r^{\sigma_{q, g}^f(j)}} \geq 0. 
\end{equation*}
Combing the two inequalities, we conclude the proof. 
\end{proof}

\section{Proof of Proposition~\ref{prop:upper_bound}}

\begin{proof}
We decompose the difference between $U_g(t)$ and its CIPW estimate as the bias and the concentration terms slightly different from the proof of Proposition~\ref{prop:lower_bound}, 
\begin{equation*}
U_g(t) - \hat{U}_{g, \lambda}^\textnormal{CIPW}(t) = \underbrace{\frac{1}{m} \sum_{i \in [m]}U_g\rbr{t \given Q_i} - \EE\sbr{\hat{U}_{g, \lambda}^{\textnormal{CIPW}}(t) \given \Qb}}_{\textnormal{Bias Term}}
 + \underbrace{U_g(t) - \frac{1}{m} \sum_{i \in [m]}U_g\rbr{t \given Q_i}}_{\textnormal{First Concentration Term}} +  \underbrace{\EE\sbr{\hat{U}_{g, \lambda}^{\textnormal{CIPW}}(t) \given \Qb} -  \hat{U}_{g, \lambda}^\textnormal{CIPW}(t)}_{\textnormal{Second Concentration Term}}. 
\end{equation*}
where $\Qb = \rbr{Q_i}_{i \in [m]}$, $\EE\sbr{\cdot \given \Qb}$ or $\EE\sbr{\cdot \given Q_i}$ denote taking expectation over all the randomness in $\cdot$ except that in $\Qb$ or $Q_i$, and
\[
U_g\rbr{t \given Q_i} \coloneqq \EE\sbr{\sum_{j \in [t]} R^{\sigma_{Q_i, g}^f(j)} \lgiven Q_i}. 
\]
For the second concentration term, we can similarly apply the empirical Bernstein bound (Theorem $4$ in~\cite{MaurerP09}) to the \iid\, random variables $\cbr{Z^{g, t}_i }_{i \in [m]}$ over the randomness in the rewards and the user observations to get that with probability at least $1 - \alpha/2$,
\[
\EE\sbr{\hat{U}_{g, \lambda}^{\textnormal{CIPW}}(t) \given \Qb} -  \hat{U}_{g, \lambda}^\textnormal{CIPW}(t) \leq \sqrt{\frac{2V_m\rbr{ \Zb^{g, t}} \ln(4/\alpha)}{m}} + \frac{7t\lambda \ln(4/\alpha)}{3(m - 1)}. 
\]
For the first concentration term, we can apply  Hoeffding's inequality~\cite{hoeffding1963} over the randomness in the queries. We can get that, with probability at least $1 - \alpha/2$, 
\[
U_g(t) - \frac{1}{m} \sum_{i \in [m]}U_g\rbr{t \given Q_i} \leq t\sqrt{\frac{\ln(2/\alpha)}{2m}}. 
\]
For the bias term, we upper bound it as follows
\begin{align*}
\frac{1}{m} \sum_{i \in [m]}U_g\rbr{t \given Q_i} - \EE\sbr{\hat{U}_{g, \lambda}^{\textnormal{CIPW}}(t) \given \Qb}  & = \frac{1}{m}\sum_{i \in [m]} \EE\sbr{\sum_{j \in [t]} R_i^{\sigma_{Q_i, g}^f(j)} - \min\rbr{\lambda p_i^{\sigma_{Q_i, g}^f(j)}, 1} R_i^{\sigma_{Q_i, g}^f(j)} \lgiven Q_i} \\ &=  \frac{1}{m}\sum_{i \in [m]} \EE\sbr{\sum_{j \in [t]}  \max\rbr{0, 1 - \lambda p_i^{\sigma_{Q_i, g}^f(j)}} R_i^{\sigma_{Q_i, g}^f(j)} \lgiven Q_i} \\
&\leq \frac{1}{m} \sum_{i \in [m]}\sum_{j \in [t]}\max\rbr{0, 1 - \lambda p_i^{\sigma_{Q_i, g}^f(j)}}. 
\end{align*}
Combining the three inequalities by applying the union bound concludes the proof. 
\end{proof}

\section{Proof of Proposition~\ref{prop:near_optimality_num_relevant}}
\begin{proof}
We only prove the proposition for $\hat{t}_g^{\textnormal{mono}}$, since the proof for $\hat{t}_g^{\textnormal{union}}$ is almost the same. 

Since $p_i^j \geq \gamma > 0$, we have $\lim_{m \rightarrow + \infty}\lambda = \lim_{m \rightarrow + \infty} \sqrt{m} > \frac{1}{\gamma} \geq \frac{1}{p_i^j}$. Thus the CIPW estimator becomes the vanilla inverse propensity weighted estimator~\cite{horvitz1952generalization} asymptotically, which is unbiased. As a result, 
\[
\lim_{m \rightarrow + \infty} \hat{U}_{g, \lambda}^{\textnormal{CIPW}}(t) = U_g(t) \quad \forall t \in \sbr{t_g^{\textnormal{max}} }
\]
holds almost surely by the strong law of large numbers.

On the other hand, $p_i^j \geq \gamma > 0$ also implies that $Z_i^{g, t} \leq \frac{t}{\gamma}$. Thus the sample variance can be bounded by $V_m\rbr{\Zb^{g, t}} \leq \frac{t^2}{\gamma^2}$ . 
Therefore,  it holds almost surely that
\[
\lim_{m \rightarrow + \infty } \rbr{U_g\rbr{\hat{t}_g^{\textnormal{mono}} - 1} -  \hat{U}_g^-\rbr{\hat{t}_g^{\textnormal{mono}} - 1, \alpha}} = 0,
\]
and 
\[
\lim_{m \rightarrow + \infty } \rbr{U_g\rbr{\hat{t}_g^{\textnormal{mono}}} -  \hat{U}_g^-\rbr{\hat{t}_g^{\textnormal{mono}}, \alpha}} = 0, 
\]
since all the terms in the differences are sub-constant in $m$. By the definition of limit at infinity, it holds almost surely that, for any $\delta > 0$, there exists $c > 0$ such that for any $m > c$, 
\[
U_g\rbr{\hat{t}_g^{\textnormal{mono}}}+ \delta > \hat{U}_g^-\rbr{\hat{t}_g^{\textnormal{mono}}, \alpha} \geq U_g^\star,
\]
and 
\[
U_g\rbr{\hat{t}_g^{\textnormal{mono}} - 1} - \delta < \hat{U}_g^-\rbr{\hat{t}_g^{\textnormal{mono}} - 1, \alpha} < U_g^\star. 
\]
From the above, we can get
\[
U_g\rbr{\hat{t}_g^{\textnormal{mono}}} \leq 1 + U_g\rbr{\hat{t}_g^{\textnormal{mono}} - 1} < 1 + U_g^\star + \delta. 
\]
\end{proof}

\section{Proof of Proposition~\ref{prop:near_optimality_threshold}}

\begin{proof}
We still only prove the proposition for $\hat{t}_g^{\textnormal{mono}}$, since the proof for $\hat{t}_{g}^{\textnormal{union}}$ is almost the same. 

Let $\delta = \min \rbr{U_g^\star - U_g\rbr{t_g^\star - 1}, U_g\rbr{t_g^\star + 1} - U_g^\star}$. Since $U_g$ is strictly increasing, we know that $\delta >0$. 
From the proof of Proposition~\ref{prop:near_optimality_num_relevant}, we know that it holds almost surely that there exists $c > 0$ such that for any $m > c$, 
\[
U_g\rbr{\hat{t}_g^{\textnormal{mono}}} >   U_g^\star - \delta \geq U_g\rbr{t_g^\star - 1}, 
\]
and thus $\hat{t}_g^{\textnormal{mono}} \geq t_g^\star $; and
\[
U_g\rbr{\hat{t}_g^{\textnormal{mono}} - 1} < U_g^\star + \delta \leq U_g\rbr{t_g^\star + 1},
\]
and thus $\hat{t}_g^{\textnormal{mono}} \leq t^\star_g + 1$. 
\end{proof}

\end{document}